\documentclass[12pt,doublespacing]{article}

\usepackage{fullpage}
\usepackage{setspace}
\usepackage{authblk}

\usepackage{cite}
\usepackage{hyperref} 

\usepackage{graphicx}

\usepackage{amsmath}
\usepackage{amssymb}
\allowdisplaybreaks

\usepackage{float}
\usepackage{subfigure}
\usepackage{makecell}
\usepackage{rotating}

\usepackage{enumitem}

\begin{document}

\title{Characteristic Mode Analysis of Plasmonic Nanostructures Using Hydrodynamic Volume Integral Equation}

\author[1]{Meruyert Khamitova}
\author[2]{Ran Zhao}
\author[3]{Doolos Aibek Uulu}
\author[1]{Sebastian Celis Sierra}
\author[1]{Hakan Bagci}

\affil[1]{Electrical and Computer Engineering (ECE) Program
\authorcr Computer, Electrical, and Mathematical Science and Engineering (CEMSE) Division
\authorcr King Abdullah University of Science and Technology (KAUST)
\authorcr Thuwal 23955-6900, Saudi Arabia
\authorcr e-mail: meruyert.khamitova@kaust.edu.sa\vspace{0.5cm}}

\affil[2]{School of Electronic Science and Engineering\
\authorcr University of Electronic Science and Technology of China (UESTC)
\authorcr Chengdu 611731, China\vspace{0.5cm}}

\affil[3]{Department of Cyber Security, Light Academy College of Engineering
\authorcr Bishkek, Kyrgyzstan}

\date{}
\maketitle
\newpage

\begin{abstract}
Metallic nanostructures confine electromagnetic fields at subwavelength scales, making them attractive as plasmonic nanoantennas. At these scales, the response of metals becomes nonlocal, and the hydrodynamic model is widely used to capture this response. However, existing solvers provide only the response to a prescribed excitation and do not directly reveal the intrinsic resonances of the structure. This work extends the characteristic mode analysis to plasmonic nanostructures to enable excitation-independent modal analysis of their resonant behavior. For simple metals, the coupled hydrodynamic and volume integral equations are reduced to a single hydrodynamic volume integral equation in terms of the induced current. The equation is discretized and cast as a generalized eigenvalue problem within the characteristic mode analysis framework, whose solution yields the characteristic mode currents and modal significance curves of the structure. The proposed framework is validated through three metallic nanostructures: a nanosphere, a nanorod, and a nanodimer. The results show that the method identifies the intrinsic resonances of each structure, including resonances not excited by a given source and additional resonances arising from the nonlocal response, which are absent in local models. The proposed framework provides physical insight into the modal mechanisms of plasmonic nanostructures and serves as a practical tool for their analysis and design.
\par\medskip
{\bf Keywords:} Characteristic mode analysis, hydrodynamic model, modal significance, nanoantennas, nonlocality, plasmonics, volume integral equation
\end{abstract}

\section{Introduction}\label{sec:intro}
Plasmonic nanostructures have gained increasing attention due to their ability to strongly confine electromagnetic fields at subwavelength scales. This confinement is enabled by surface plasmons, which are collective oscillations of free electrons excited at metal surfaces~\cite{maier2007}. This strong field localization is exploited in many applications, including near-field scanning optical microscopy~\cite{schuller2010}, biomedical sensing and detection~\cite{stockman2015}, and thin-film photovoltaic enhancement~\cite{atwater2010}. Due to their enhanced scattering and absorption characteristics, metallic nanostructures are widely used as plasmonic nanoantennas to manipulate light~\cite{yla-oijala2017, opticalantennas2011}. The performance of these nanoantennas is governed by their resonant behavior, which is determined by their geometry and material composition. Accurate simulation tools are therefore essential for tuning these parameters to achieve the desired electromagnetic response.

The optical response of a metal is usually well described by the local-response approximation, typically using Drude or Drude--Lorentz models~\cite{schmitt2018}. However, as the dimensions of the structure approach the nanoscale, spatial dispersion effects become significant, and this approximation may no longer provide an accurate description~\cite{mortensen2012}. In this regime, the collective motion of free electrons can support longitudinal plasma waves inside the metal, which are absent in the local model~\cite{mortensen2013}. Consequently, the metal permittivity must be expressed as a nonlocal function, $\varepsilon(\mathbf{r},\mathbf{r}^{\prime})$, which depends on both the observation point $\mathbf{r}$ and the source point $\mathbf{r}^{\prime}$, capturing spatial dispersion~\cite{zheng2018}.

While fully quantum simulations capture nanoscale electronic effects, including nonlocal response, electron spill-out, and quantum tunneling, their high computational cost restricts them to structures of only a few nanometers~\cite{esteban2012bridging, teperik2013}. In the deep-nanometer regime, however, nonlocality is often the dominant quantum correction, whereas other quantum effects are less pronounced. As a result, the hydrodynamic model has become a widely adopted semiclassical approach for describing the nonlocal response~\cite{schmitt2018,zheng2018}. 
In this model, the free-electron gas is treated as a continuous charge fluid, whose motion is described by the hydrodynamic equation (HDE), which extends the Drude model to include nonlocality. Coupling the HDE with the Maxwell equations enables the analysis of nonlocal electromagnetic effects in metallic nanostructures~\cite{kupresak2018comparison, doolos2023, schmitt2018, zheng2018, fang2017, gungor2022, bhardwaj2018, hiremath2012, ciraci2013}. A recently developed method couples the volume integral equation (VIE) with the HDE to analyze scattering from composite metal--dielectric nanostructures near the plasma frequency~\cite{doolos2023}. However, because this approach provides the response to a prescribed excitation, it does not directly reveal the intrinsic resonant properties of the structure, which are essential for systematic nanoantenna design.

The characteristic mode analysis (CMA) yields a set of intrinsic modes that depend only on the geometry and the material of the structure~\cite{garbacz1965}. These modes form a basis for representing the induced current under arbitrary excitation. Because they are independent of the excitation, the CMA provides direct access to the intrinsic resonances of a structure~\cite{chen2015}. Mathematically, the CMA is formulated as an eigenvalue problem constructed from the discretized form of the governing integral equation. The resulting eigenvectors represent the characteristic modes of the structure, and the corresponding eigenvalues quantify their resonant behavior. The CMA has been widely applied in the analysis of various nanostructures using integral equation methods~\cite{lau2022}. In particular, its integration with the surface integral equations has been extensively studied~\cite{harrington1971,chen2016,huang2023,hu2016,yla-oijala2017,ran2022}. However, this approach may produce spurious modes and does not explicitly distinguish between absorption, radiation, and extinction resonances~\cite{lau2022}. The CMA has also been combined with the VIE to analyze dielectric and magnetic bodies at microwave frequencies~\cite{harrington1972,kuosmanen2022,wu2017}. Despite its accuracy in the microwave regime, this formulation relies on a local model and therefore is not directly suitable for metallic nanostructures operating near the plasma frequency.

This work integrates the coupled system of the HDE and the VIE with the CMA to enable excitation-independent modal analysis and thereby characterize the resonant behavior of plasmonic nanostructures. The analysis is restricted to simple (alkali) metals, which are modeled as containing only free electrons~\cite{forstmann1986metal}. The motion of these electrons gives rise to a hydrodynamic current density induced inside the metal. The coupled system of the HDE and the VIE reduces to a single equation in terms of this hydrodynamic current density alone, which is termed the hydrodynamic volume integral equation (HDVIE). To discretize the HDVIE, the scatterer is partitioned into a mesh of tetrahedral elements and the hydrodynamic current is expanded using full Schaubert--Wilton--Glisson (SWG) basis functions~\cite{schaubert1984}; half SWG functions are excluded to enforce the additional boundary condition required by the HDE at the interface between the metal and the background medium~\cite{doolos2023}. Galerkin testing then yields the discretized form of the HDVIE as a matrix system, which is used to construct the generalized eigenvalue equation (GEE) within the CMA framework. Solving this eigenvalue problem over the frequency range of interest yields the characteristic hydrodynamic currents and their modal significance (MS) curves, providing excitation-independent access to the resonant modes of the structure. A preliminary version of this HDVIE--CMA framework is reported in~\cite{khamitova2024}.

The main contributions of this work are threefold. First, the coupled system of the VIE and the HDE is reduced to the single HDVIE in terms of the hydrodynamic current density under the assumption of simple (alkali) metals. Second, the GEE is formulated from the discretized HDVIE within the CMA framework, enabling excitation-independent analysis of the resonant behavior of metallic nanostructures. Third, the proposed HDVIE--CMA formulation is assessed through three numerical examples by comparing the identified resonances with extinction spectra under plane wave and dipole excitations and, where applicable, analytical Mie-series solutions.

The remainder of this paper is organized as follows. Section~\ref{sec:formulation} presents the formulation, including the derivation of the HDVIE and its discretization, and the construction of the CMA framework. Section~\ref{sec:num_results} provides several numerical examples to verify the accuracy and reliability of the proposed method. Finally, Section~\ref{sec:conclusion} summarizes the main findings of this work.

\section{Formulation}\label{sec:formulation}
\subsection{HDVIE}\label{sec:hdvie_formulation}

\begin{figure}[ht!]
\centering
\includegraphics[width=0.4\columnwidth]{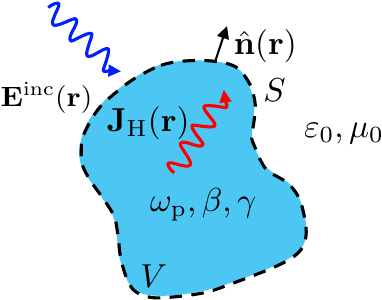}
  \caption{Description of the electromagnetic scattering problem.}\label{problem_desc}
\end{figure}

Consider an arbitrarily shaped, nonmagnetic metallic nanostructure occupying volume $V$ with boundary surface $S$ (Fig.~\ref{problem_desc}). The nanostructure is embedded in an unbounded homogeneous background medium characterized by permeability $\mu_0$ and permittivity $\varepsilon_0$. The structure is illuminated by a time-harmonic incident electric field $\mathbf{E}^{\mathrm{inc}}(\mathbf{r})$, where an $e^{\mathrm{j}\omega t}$ time dependence is assumed and suppressed. Here, $\omega$ is the angular frequency. The free electrons respond collectively to this excitation, giving rise to an induced hydrodynamic current density $\mathbf{J}_{\mathrm{H}}(\mathbf{r})$ within $V$~\cite{doolos2023,zheng2018}. For the simple (alkali) metals considered here, the bound-charge response is negligible, so $\mathbf{J}_{\mathrm{H}}(\mathbf{r})$ captures the entire induced response. 

The induced current $\mathbf{J}_{\mathrm{H}}(\mathbf{r})$ generates a scattered field $\mathbf{E}^{\mathrm{sca}}(\mathbf{r})$. Using the volume equivalence principle~\cite{jin2015}, $\mathbf{E}^{\mathrm{sca}}(\mathbf{r})$ is expressed as
\begin{equation}
    \mathbf{E}^{\mathrm{sca}}(\mathbf{r}) =\mathcal{L}[\mathbf{J}_\mathrm{H}](\mathbf{r}).
    \label{eq:sca_field}
\end{equation}
Here, the volume integral operator $\mathcal{L}[\mathbf{X}](\mathbf{r})$ is defined as~\cite{jin2015}
\begin{equation}
\mathcal{L}[\mathbf{X}](\mathbf{r})= -\mathrm{j} \omega \mu_0 \int_V \mathbf{X}(\mathbf{r}^{\prime}) G_0(\mathbf{r}, \mathbf{r}^{\prime})\,d v^{\prime}+\frac{1}{\mathrm{j} \omega \varepsilon_0} \nabla\int_V \nabla^\prime \cdot \mathbf{X}(\mathbf{r}^{\prime}) G_0(\mathbf{r}, \mathbf{r}^{\prime})\,d v^{\prime}
\label{eq:l_operator}
\end{equation}
where $G_0(\mathbf{r}, \mathbf{r}^{\prime})=e^{-\mathrm{j} k_0 R} /(4 \pi R)$ is the Green function of the background medium, $k_0=\omega\sqrt{\varepsilon_0 \mu_0}$ is the corresponding wavenumber, and $R=|\mathbf{r}-\mathbf{r}^{\prime}|$ is the distance between points $\mathbf{r}$ and $\mathbf{r}^\prime$.

The total electric field $\mathbf{E}(\mathbf{r})$ is the superposition of the incident and scattered fields:
\begin{equation}
    \mathbf{E}(\mathbf{r}) =\mathbf{E}^{\mathrm{inc}}(\mathbf{r}) + \mathbf{E}^{\mathrm{sca}}(\mathbf{r}).
    \label{eq:efield}
\end{equation}
The collective response of the free electrons to $\mathbf{E}(\mathbf{r})$ is described by the free-electron hydrodynamic model~\cite{forstmann1986metal,raza2011unusual}:
\begin{equation}
    (\mathrm{j}\omega +\gamma) \mathbf{J}_{\mathrm{H}}(\mathbf{r})+\beta^2 \nabla \rho_\mathrm{H}(\mathbf{r})=\omega_{\mathrm{p}}^2 \varepsilon_0 \mathbf{E}(\mathbf{r}), \quad \mathbf{r} \in V.
    \label{eq:hde}
\end{equation}
Here, $\omega_{\mathrm{p}}$ is the plasma frequency, $\gamma$ is the damping constant, $\beta$ is the nonlocal parameter, given by $\beta^2 = 0.6\,v_{\mathrm{F}}^2$, where $v_{\mathrm{F}}$ is the Fermi velocity, and $\rho_{\mathrm{H}}(\mathbf{r})$ is the free-electron charge density. The charge density is related to $\mathbf{J}_{\mathrm{H}}(\mathbf{r})$ through the continuity equation:
\begin{equation}
\nabla \cdot \mathbf{J}_{\mathrm{H}}(\mathbf{r}) = -\mathrm{j}\omega \rho_\mathrm{H}(\mathbf{r}), \quad \mathbf{r} \in V.
\label{eq:current_charge}
\end{equation}
Equation~\eqref{eq:hde} is referred to as the HDE, which is an extension of the Drude model that includes the gradient term $\beta^2 \nabla \rho_\mathrm{H}(\mathbf{r})$. This term accounts for electron-gas pressure, captures the nonlocal electromagnetic response, and gives rise to longitudinal wave propagation inside the nanostructure~\cite{raza2015}. Because this nonlocal model supports an additional longitudinal wave, the standard Maxwell boundary conditions alone are not sufficient. Consequently, the HDE~\eqref{eq:hde} requires an additional boundary condition~\cite{boardman1981} on the boundary surface $S$ of $V$:
\begin{equation}
    \hat{\mathbf{n}}(\mathbf{r}) \cdot \mathbf{J}_{\mathrm{H}}(\mathbf{r}) = 0,\quad \mathbf{r} \in S
    \label{eq:bc}
\end{equation}
where $\hat{\mathbf{n}}(\mathbf{r})$ denotes the outward-pointing unit normal vector on $S$. This boundary condition is known as the hard-wall condition, which confines the electron motion to the metallic region (i.e., electron spill-out is neglected). Because the electrons cannot cross the boundary, the normal component of $\mathbf{J}_{\mathrm{H}}(\mathbf{r})$ vanishes at the interface between the metal and the background medium. 

Substituting~\eqref{eq:sca_field},~\eqref{eq:efield}, and~\eqref{eq:current_charge} into~\eqref{eq:hde} yields the HDVIE for the unknown current density $\mathbf{J}_{\mathrm{H}}(\mathbf{r})$:
\begin{equation}
\frac{\mathrm{j} \beta^2}{\omega \omega_{\mathrm{p}}^2 \varepsilon_0} \nabla\left[\nabla \cdot \mathbf{J}_{\mathrm{H}}(\mathbf{r})\right] +\frac{\mathrm{j} \omega + \gamma}{ \omega_{\mathrm{p}}^2 \varepsilon_0} \mathbf{J}_{\mathrm{H}}(\mathbf{r}) -
\mathcal{L}[\mathbf{J}_{\mathrm{H}}](\mathbf{r}) =\mathbf{E}^{\mathrm{inc}}(\mathbf{r}), \quad \mathbf{r} \in V. 
\label{eq:hdvie}
\end{equation}
This equation is discretized using the scheme described in Section~\ref{sec:hdvie_discretization}, which also enforces the boundary condition in~\eqref{eq:bc}.

\subsection{Discretization of HDVIE}\label{sec:hdvie_discretization}
To discretize~\eqref{eq:hdvie}, volume $V$ is partitioned into a mesh of tetrahedral elements, and $\mathbf{J}_\mathrm{H}(\mathbf{r})$ is then expanded using SWG basis functions~\cite{schaubert1984} as
\begin{equation}
    \mathbf{J}_{\mathrm{H}} (\mathbf{r}) = \sum_{n=1}^{N} \{ \bar{I}\}_n \mathbf{f}_n(\mathbf{r}),\quad \mathbf{r} \in {V}
    \label{eq:expansion}
\end{equation}
where $\{\bar{I}\}_n$ are the unknown expansion coefficients, $\mathbf{f}_n(\mathbf{r})$ are SWG basis functions associated with the $n$th triangle $S_n$ of the tetrahedral mesh, and $N$ is the total number of basis functions. The SWG basis function $\mathbf{f}_n(\mathbf{r})$ is defined as follows:
\begin{equation}
    \mathbf{f}_n(\mathbf{r})= \begin{cases}\mathbf{f}_n^{+}(\mathbf{r})=\frac{|S_n|}{3|V_n^{+}|}(\mathbf{r}-\mathbf{r}_n^{+}), & \mathbf{r} \in V_n^{+} \\ \mathbf{f}_n^{-}(\mathbf{r})=-\frac{|S_n|}{3|V_n^{-}|}(\mathbf{r}-\mathbf{r}_n^{-}), & \mathbf{r} \in V_n^{-} \\ 0, & \mathrm { elsewhere }\end{cases}.
    \label{eq:swg}
\end{equation}
Here, $V_n^{+}$ and $V_n^{-}$ are the tetrahedra that share $S_n$ on opposite sides, $\mathbf{r}_n^\pm$ are the free nodes of $V_n^\pm$, $|S_n|$ is the area of $S_n$, and $|V_n^\pm|$ are the volumes of $V_n^\pm$. The use of full SWG basis functions, defined on pairs of adjacent tetrahedra, enforces continuity of the normal component of $\mathbf{J}_\mathrm{H}(\mathbf{r})$ across internal triangles. Moreover, half SWG basis functions $\mathbf{f}_n^{\pm}(\mathbf{r})$ associated with boundary triangles on $S$ are excluded from the basis set. This choice is consistent with the boundary condition in~\eqref{eq:bc}, which requires the normal component of $\mathbf{J}_\mathrm{H}(\mathbf{r})$ to vanish on $S$~\cite{doolos2023}.

Substituting~\eqref{eq:expansion} into~\eqref{eq:hdvie} and applying Galerkin testing with $\mathbf{f}_m(\mathbf{r})$, $m=1,2,\ldots,N$, yields the $N\times N$ matrix system:
\begin{equation}
    \bar{\bar{Z}} \bar{I} = \bar{V}^{\mathrm{inc}}.
    \label{eq:matrix_system}
\end{equation}
In~\eqref{eq:matrix_system}, the entries of the matrix $\bar{\bar{Z}}$ are given by
\begin{equation}
\begin{aligned} 
\{\bar{\bar{Z}}\}_{mn} &= \frac{\mathrm{j} \omega+\gamma}{\omega_{\mathrm{p}}^2 \varepsilon_0}\int_{V_m\cap V_n}\mathbf{f}_m(\mathbf{r})\cdot\mathbf{f}_n(\mathbf{r})\,dv-\frac{\mathrm{j} \beta^2}{\omega\omega_{\mathrm{p}}^2 \varepsilon_0}\int_{V_m\cap V_n}[\nabla\cdot\mathbf{f}_m(\mathbf{r})][\nabla\cdot\mathbf{f}_n(\mathbf{r})]\,dv \\ 
+&\mathrm{j}\omega\mu_0 \int_{V_m}\int_{V_n}\mathbf{f}_m(\mathbf{r})\cdot\mathbf{f}_n(\mathbf{r}^{\prime})\,G_0(\mathbf{r},\mathbf{r}^{\prime})\,dv^{\prime}dv \\ +& \frac{1}{\mathrm{j}\omega\varepsilon_0} \int_{V_m}\int_{V_n}[\nabla\cdot\mathbf{f}_m(\mathbf{r})][\nabla^{\prime}\cdot\mathbf{f}_n(\mathbf{r}^{\prime})]G_0(\mathbf{r},\mathbf{r}^{\prime})\,dv^{\prime}\,dv
\end{aligned} \label{eq:matrix_entries}
\end{equation}
for $m,n=1,2,\ldots,N$. Here, $V_m = V_m^{+}\cup V_m^{-}$ denotes the support of $\mathbf{f}_m(\mathbf{r})$, and similarly for $V_n$. The second and fourth terms in~\eqref{eq:matrix_entries} follow from integration by parts. The boundary terms vanish because each $\mathbf{f}_m(\mathbf{r})$ has vanishing normal component on all faces of $V_m$ except $S_m$, across which the normal component is continuous.

The entries of the right-hand side vector $\bar{V}^{\mathrm{inc}}$ are given by
\begin{equation}
    \label{eq:rhsd}
    \{ \bar{V}^{\mathrm{inc}} \}_{m}= \int_{V_m} \mathbf{f}_m(\mathbf{r})\cdot \mathbf{E}^{\mathrm{inc}}(\mathbf{r}) \,dv
\end{equation}
for $m=1,2,\ldots,N$.

\subsection{CMA using HDVIE}\label{sec:cma_section}
In the CMA, the GEE~\cite{harrington1971} is expressed as 
\begin{equation}
    \bar{\bar{Z}}\bar{I}_k=(1+\mathrm{j} \lambda_k) \bar{\bar{W}}\bar{I}_k
    \label{eq:gee}
\end{equation}
where $\bar{I}_k$ denotes the $k$th characteristic current mode, and $\lambda_k$ is the corresponding characteristic eigenvalue. Here, $\bar{\bar{Z}}$ is the impedance matrix obtained from the discretization of the HDVIE~\eqref{eq:hdvie}, given in~\eqref{eq:matrix_system}, and $\bar{\bar{W}}$ is a weighting matrix that determines which resonance (extinction, radiation, or
absorption) the GEE characterizes~\cite{kuosmanen2022}. In this work, extinction resonances, which occur when the reactive power becomes small relative to the extinction power, are considered. Accordingly, the weighting matrix is chosen as $\bar{\bar{W}}=\bar{\bar{R}}$, where $\bar{\bar{R}}=\mathrm{Re}\{\bar{\bar{Z}}\}$ is associated with the extinction power (see Section~\ref{sec:physical_interp}). Writing the impedance matrix as $\bar{\bar{Z}} = \bar{\bar{R}} + \mathrm{j}\bar{\bar{X}}$, with $\bar{\bar{X}}=\mathrm{Im}\{\bar{\bar{Z}}\}$, and substituting $\bar{\bar{W}}=\bar{\bar{R}}$ into~\eqref{eq:gee} gives
\begin{equation}
    (\bar{\bar{R}} + \mathrm{j}\bar{\bar{X}})\bar{I}_k
    = (1+\mathrm{j}\lambda_k)\bar{\bar{R}}\bar{I}_k.
    \label{eq:gee_expanded}
\end{equation}
Canceling the common term $\bar{\bar{R}}\bar{I}_k$ from both sides reduces this to
\begin{equation}
    \bar{\bar{X}}\bar{I}_k=\lambda_k \bar{\bar{R}} \bar{I}_k.
    \label{eq:gee2}
\end{equation}
Under the Galerkin discretization, both $\bar{\bar{R}}$ and $\bar{\bar{X}}$ are real and symmetric. Moreover, $\bar{\bar{R}}$ is positive definite, and thus the resulting characteristic modes and eigenvalues are real. The solution procedure for the reduced GEE in~\eqref{eq:gee2} is described in Section~\ref{sec:numerical_sol_gee}. These modes are then normalized with respect to the weighting matrix $\bar{\bar{R}}$, leading to the orthogonality relation~\cite{harrington1971}
\begin{equation}
     \langle \bar{I}_j, \bar{\bar{R}} \bar{I}_k \rangle = \delta_{jk}
    \label{eq:orthogonality}
\end{equation}
where $\langle \bar{u}, \bar{v} \rangle = \bar{u}^{\top} \bar{v}$ denotes the symmetric bilinear product of two vectors $\bar{u}$ and $\bar{v}$, and $\delta_{jk}$ is the Kronecker delta. This property enables the expansion of the current in terms of the characteristic modes as
\begin{equation}
    \bar{I}=\sum_{k=1}^N a_k \bar{I}_k.
    \label{eq:current}
\end{equation}
Here, $a_k$ is the expansion coefficient associated with the characteristic mode $\bar{I}_k$. Inserting~\eqref{eq:current} into~\eqref{eq:matrix_system} and using~\eqref{eq:gee} with $\bar{\bar{W}}=\bar{\bar{R}}$ yields
\begin{equation}
    \sum_{k=1}^N a_k (1+\mathrm{j} \lambda_k) \bar{\bar{R}}\bar{I}_k = \bar{V}^\mathrm{inc}.
    \label{eq:matrix_aux}
\end{equation}
To obtain the expansion coefficients, both sides of~\eqref{eq:matrix_aux} are multiplied by $\bar{I}_j^{\top}$, forming the product $\langle \bar{I}_j, \cdot \rangle$. Using the orthogonality relation~\eqref{eq:orthogonality}, the Kronecker delta $\delta_{jk}$ collapses the sum to the single term $k=j$, giving 
\begin{equation}
    a_j (1+\mathrm{j}\lambda_j)= \langle \bar{I}_j, \bar{V}^{\mathrm{inc}} \rangle = V_j.
    \label{eq:aux_coef}
\end{equation}
Here, $V_j=\bar{I}_j^{\top} \bar{V}^{\mathrm{inc}}$ is termed the modal excitation coefficient, and it measures the strength of the coupling between the incident electric field $\mathbf{E}^{\mathrm{inc}}(\mathbf{r})$ and the $j$th characteristic mode with current $\bar{I}_j$. Solving~\eqref{eq:aux_coef} for the coefficient $a_j$ yields
\begin{equation}
    a_j = \frac{V_j}{1+\mathrm{j}\lambda_j}.
    \label{eq:exp_coef}
\end{equation}
From~\eqref{eq:exp_coef}, the excitation-independent parameter, known as the MS~\cite{chen2015}, can be derived as
\begin{equation}
     \sigma_j= \frac{1}{|1+\mathrm{j}\lambda_j|}.
     \label{eq:ms}
\end{equation}
This parameter quantifies the intrinsic resonant strength of each mode at a given frequency. Together with the modal excitation coefficient, $\sigma_j$ determines the contribution of the $j$th mode to the total electromagnetic response under a given excitation~\cite{chen2015}. The value of $\sigma_j$ ranges from $0$ to $1$, where modes whose $\sigma_j$ approaches $1$ represent the naturally resonating modes supported by the structure. Unlike the characteristic eigenvalue $\lambda_j$, which can take any value from $-\infty$ to $+\infty$, $\sigma_j$ provides a more convenient measure for evaluating the resonant behavior of modes over a broad frequency spectrum. A more detailed physical interpretation of the eigenvalues, and hence of the MS, is provided in Section~\ref{sec:physical_interp}. 

\subsection{Physical Interpretation}\label{sec:physical_interp}
The physical meaning of the eigenvalues, and of the weighting-matrix choice $\bar{\bar{W}}=\bar{\bar{R}}$  made in Section~\ref{sec:cma_section}, follows from the complex power balance expressed by the Poynting theorem~\cite{chen2015}. The complex power $P$ delivered by the incident field $\mathbf{E}^{\mathrm{inc}}(\mathbf{r})$ to the induced free-electron current density $\mathbf{J}_{\mathrm{H}}(\mathbf{r})$ is
\begin{equation}
    P= \frac{1}{2} \int_V \mathbf{E}^{\mathrm{inc}}(\mathbf{r})\cdot \mathbf{J}_\mathrm{H}^*(\mathbf{r}) \,d v
    \label{eq:inc_power}
\end{equation}
where the superscript $*$ denotes complex conjugation. Applying the complex Poynting theorem to the scattered fields $\mathbf{E}^{\mathrm{sca}}(\mathbf{r})$ and $\mathbf{H}^{\mathrm{sca}}(\mathbf{r})$ generated by $\mathbf{J}_{\mathrm{H}}(\mathbf{r})$ gives
\begin{equation}
    \begin{aligned}
    \frac{1}{2}\int_V \mathbf{E}^{\mathrm{sca}}(\mathbf{r})\cdot\mathbf{J}_\mathrm{H}^*(\mathbf{r})\,dv&= -\frac{1}{2}\oint_S[\mathbf{E}^{\mathrm{sca}}(\mathbf{r})\times \mathbf{H}^{\mathrm{sca}*}(\mathbf{r})]\cdot\hat{\mathbf{n}}(\mathbf{r})\,ds \\
    & -\frac{\mathrm{j}\omega}{2}\int_V[\mu_0|\mathbf{H}^{\mathrm{sca}}(\mathbf{r})|^2- \varepsilon_0|\mathbf{E}^{\mathrm{sca}}(\mathbf{r})|^2]\,dv.
    \end{aligned}
    \label{eq:poynting_sca}
\end{equation}
Substituting $\mathbf{E}^{\mathrm{sca}}(\mathbf{r})=\mathbf{E}(\mathbf{r})-\mathbf{E}^{\mathrm{inc}}(\mathbf{r})$ [see~\eqref{eq:efield}] into the left-hand side of~\eqref{eq:poynting_sca} and using~\eqref{eq:inc_power}, the complex power $P$ is expressed as
\begin{equation}
    \begin{aligned}
    P ={} & \frac{1}{2}\oint_S[\mathbf{E}^{\mathrm{sca}}(\mathbf{r})\times \mathbf{H}^{\mathrm{sca}*}(\mathbf{r})]\cdot\hat{\mathbf{n}}(\mathbf{r})\,ds +\frac{\mathrm{j}\omega}{2}\int_V[\mu_0|\mathbf{H}^{\mathrm{sca}}(\mathbf{r})|^2- \varepsilon_0|\mathbf{E}^{\mathrm{sca}}(\mathbf{r})|^2]\,dv \\
    & +\frac{1}{2}\int_V \mathbf{J}_\mathrm{H}^*(\mathbf{r})\cdot\mathbf{E}(\mathbf{r})\,dv.
    \end{aligned}
    \label{eq:power_theorem}
\end{equation}
The surface integral is evaluated over the boundary $S$ of $V$. Because $\mathbf{J}_\mathrm{H}(\mathbf{r})$ is confined to $V$, no sources lie outside $S$, consequently, the real part of this integral gives the power scattered to the far field while its imaginary part contributes to the reactive power. Using the HDE~\eqref{eq:hde}, the last term of~\eqref{eq:power_theorem} is expanded as
\begin{equation}
\begin{aligned} \int_V \mathbf{J}^*_\mathrm{H}(\mathbf{r})\cdot\mathbf{E}(\mathbf{r})\,dv& =\frac{\mathrm{j}\omega}{\omega_\mathrm{p}^2\varepsilon_0} \int_V|\mathbf{J}_\mathrm{H}(\mathbf{r})|^2\,dv - \frac{\mathrm{j}\omega\beta^2}{\omega_\mathrm{p}^2\varepsilon_0} \int_V|\rho_\mathrm{H}(\mathbf{r})|^2\,dv +\frac{\gamma}{\omega_\mathrm{p}^2\varepsilon_0} \int_V|\mathbf{J}_\mathrm{H}(\mathbf{r})|^2\,dv \\
& + \underbrace{\frac{\beta^2}{\omega_\mathrm{p}^2\varepsilon_0}\int_V\nabla\cdot[\rho_\mathrm{H}(\mathbf{r}) \mathbf{J}^*_\mathrm{H}(\mathbf{r})]dv}_{Q} 
\end{aligned}
\label{eq:hydro_power} 
\end{equation}    
Applying the divergence theorem, $Q$ is converted into a surface flux which vanishes by the boundary condition in~\eqref{eq:bc}:
\begin{equation}
Q = \frac{\beta^2}{\omega_\mathrm{p}^2 \varepsilon_0} \oint_S\rho_\mathrm{H}(\mathbf{r})[\mathbf{J}^*_\mathrm{H}(\mathbf{r})\cdot\hat{\mathbf{n}}(\mathbf{r})]\,ds = 0.
\label{eq:Q_flux}
\end{equation}
Substituting~\eqref{eq:hydro_power} into~\eqref{eq:power_theorem} and separating real and imaginary parts yields
\begin{equation}
    P = P^{\mathrm{ext}} + \mathrm{j} P^{\mathrm{reac}}
    \label{eq:power_decomp}
\end{equation}
where $P^{\mathrm{ext}} = P^{\mathrm{sca}} + P^{\mathrm{abs}}$ is the extinction power, $P^{\mathrm{reac}}$ is the reactive power, $P^{\mathrm{sca}}$ is the scattered power, and $P^{\mathrm{abs}}$ is the absorbed power~\cite{polimeridis2015}. Expressions of $P^{\mathrm{sca}}$, $P^{\mathrm{abs}}$, and $P^{\mathrm{reac}}$ read
\begin{equation}
\begin{aligned}
    P^{\mathrm{sca}}& = \frac{1}{2}\,\mathrm{Re}\left\{\oint_S[\mathbf{E}^{\mathrm{sca}}(\mathbf{r}) \times\mathbf{H}^{\mathrm{sca}*}(\mathbf{r})]\cdot\hat{\mathbf{n}}(\mathbf{r})\,ds\right\}\\
    P^{\mathrm{abs}} &= \frac{\gamma}{2\omega_\mathrm{p}^2\varepsilon_0}\int_V\left|\mathbf{J}_\mathrm{H}(\mathbf{r})\right|^2 dv\\
    P^{\mathrm{reac}}& =\frac{1}{2}\,\mathrm{Im}\left\{\oint_S[\mathbf{E}^{\mathrm{sca}}(\mathbf{r}) \times\mathbf{H}^{\mathrm{sca}*}(\mathbf{r})]\cdot\hat{\mathbf{n}}(\mathbf{r})\,ds\right\}+\frac{\omega}{2}\int_V[\mu_0|\mathbf{H}^{\mathrm{sca}}(\mathbf{r})|^2- \varepsilon_0|\mathbf{E}^{\mathrm{sca}}(\mathbf{r})|^2]\,dv \\
    &+ \frac{\omega}{2}\left[E^\mathrm{kin} - E^\mathrm{pot}\right]
\end{aligned}
\label{eq:power_definitions}
\end{equation}
where 
\begin{equation}
\begin{aligned}
E^\mathrm{kin}& = \frac{1}{\omega_\mathrm{p}^2\varepsilon_0}\int_V|\mathbf{J}_\mathrm{H}(\mathbf{r})|^2\,dv\\
E^\mathrm{pot}& = \frac{\beta^2}{\omega_\mathrm{p}^2\varepsilon_0}\int_V|\rho_\mathrm{H}(\mathbf{r})|^2\,dv
\end{aligned}
\label{eq:energy_definitions}
\end{equation}
are the kinetic and potential energy contributions, respectively~\cite{forstmann1986metal}. The kinetic energy contribution $E^\mathrm{kin}$ is associated with the motion of the free electrons, while the potential energy contribution $E^\mathrm{pot}$ arises from the compression of the electron gas, captured by the nonlocal pressure term $\beta^2\nabla\rho_\mathrm{H}(\mathbf{r})$ in the HDE~\eqref{eq:hde}.

Substituting the basis expansion~\eqref{eq:expansion} into the power definition~\eqref{eq:inc_power} gives
\begin{equation}
P = \frac{1}{2}\bar{I}^\dagger \bar{V}^{\mathrm{inc}}
\label{eq:power_discrete1}
\end{equation}
where the superscript $\dagger$ denotes the conjugate transpose. Using the matrix system~\eqref{eq:matrix_system} to replace $\bar{V}^{\mathrm{inc}}$, $P$ is written in terms of the impedance matrix $\bar{\bar{Z}}$ as in~\cite{harrington1972}
\begin{equation}
    P=\frac{1}{2}\bar{I}^\dagger \bar{\bar{Z}} \bar{I}=\frac{1}{2}\bar{I}^\dagger  \bar{\bar{R}} \bar{I}+\mathrm{j}\frac{1}{2}\bar{I}^\dagger  \bar{\bar{X}} \bar{I}.
    \label{eq:power_discrete2}
\end{equation}
Comparing~\eqref{eq:power_discrete2} with the decomposition~\eqref{eq:power_decomp}, $\bar{\bar{R}}$ and $\bar{\bar{X}}$ are associated with the extinction and reactive power contributions, respectively. When $\bar{\bar{W}}=\bar{\bar{R}}$, the GEE in~\eqref{eq:gee} reduces to~\eqref{eq:gee2}. Multiplying both sides of~\eqref{eq:gee2} with $\bar{I}_k^{\top}$ forming the product $\langle \bar{I}_k, \cdot \rangle$, the expression for the corresponding eigenvalue is obtained as
\begin{equation}
    \lambda_k = \frac{\langle \bar{I}_k, \bar{\bar{X}}\bar{I}_k \rangle}
    {\langle \bar{I}_k, \bar{\bar{R}}\bar{I}_k \rangle}
    = \frac{P_k^{\mathrm{reac}}}{P_k^{\mathrm{ext}}}.
    \label{eq:eigenvalue}
\end{equation}
Here, $P_k^{\mathrm {reac}}$ and $P_k^{\mathrm {ext}}$ represent the reactive and extinction power contributions associated with the characteristic mode with current $\bar{I}_k$, respectively. Note that the complex conjugation is omitted in~\eqref{eq:eigenvalue} because $\bar{I}_k$ obtained from~\eqref{eq:gee2} are real, as explained in Section~\ref{sec:cma_section}. The eigenvalue associated with each characteristic current quantifies the ratio of reactive power to extinction power. Positive and negative eigenvalues correspond to the dominance of inductive and capacitive energies, respectively~\cite{chen2015}. An eigenvalue close to zero indicates that the extinction power dominates the reactive power, corresponding to an extinction resonance. By~\eqref{eq:ms}, such a mode has an MS approaching $1$, whereas nonresonant modes, for which the reactive power is large, have an MS closer to $0$. The MS therefore provides a bounded and convenient metric for identifying the resonant modes of the structure and evaluating their behavior across frequency.

\subsection{Numerical Solution of GEE}
\label{sec:numerical_sol_gee}
Solving~\eqref{eq:gee2} yields as many characteristic modes as there are unknowns in the discretized system.
However, the structure's response is typically governed by a few extinction-resonant modes. As shown in Section~\ref{sec:physical_interp}, the eigenvalues of these modes have the smallest magnitudes. Therefore, a Krylov-subspace iterative method is used instead of a direct eigensolver. In this work, the implicitly restarted Arnoldi method (IRAM)~\cite{cheng2018}, as implemented in the ARPACK library~\cite{lehoucq1998}, is employed to compute the desired subset of eigenpairs. Because IRAM is efficient at computing extremal (largest-magnitude) eigenvalues~\cite{lehoucq1998},~\eqref{eq:gee2} is transformed into the standard eigenvalue equation
\begin{equation}
    \bar{\bar{X}}^{-1}\bar{\bar{R}} \bar{I}_k=\tilde{\lambda}_k\bar{I}_k
    \label{eq:gee3}
\end{equation}
where $\tilde{\lambda}_k=\lambda_k^{-1}$. Under this transformation, the smallest-magnitude eigenvalues of~\eqref{eq:gee2}, corresponding to the resonant modes, become the largest-magnitude eigenvalues of~\eqref{eq:gee3}, which IRAM computes efficiently.

Each IRAM iteration requires a matrix-vector product $\bar{w} = \bar{\bar{X}}^{-1}\bar{\bar{R}}\bar{z}$, evaluated in two steps: (1)~computing $\bar{y}=\bar{\bar{R}}\bar{z}$ and (2)~solving $\bar{\bar{X}}\bar{w}=\bar{y}$ for $\bar{w}$. The second step is carried out iteratively using the transpose-free quasi-minimal residual (TFQMR) method~\cite{freund1993}. The iterations are terminated when the relative residual satisfies
\begin{equation}
    \frac{\|\bar{y}-\bar{\bar{X}} \bar{w}^{(n)}\|_2}{\|\bar{y}\|_2}<10^{-4}
\end{equation}
where $\bar{w}^{(n)}$ is the solution vector at iteration $n$ and $\|\cdot\|_2$ denotes the $\ell_2$ norm.

To analyze the broadband behavior of characteristic modes, a mode-tracking procedure is used to ensure that modes with similar current distributions are identified as the same physical mode across frequency. The mode-tracking algorithm used in this work follows the approach in~\cite{chen2015}.

\section{Numerical Results} \label{sec:num_results}
In this section, several numerical examples are presented to verify the accuracy and reliability of the proposed HDVIE-based CMA formulation. The resonances identified from the MS curves are compared with those of the normalized extinction power computed after solving the discretized HDVIE~\eqref{eq:matrix_system} under two different excitations: a plane wave and an infinitesimal dipole. These excitations are considered because they couple to different subsets of the characteristic modes and therefore different resonances appear in the extinction spectra. The extinction power, denoted by $P^\mathrm{ext}$, is computed following the procedure in~\cite{polimeridis2015} as
\begin{equation}
\begin{aligned}
    P^{\mathrm {ext}}&=\frac{1}{2} \mathrm{Re}\left\{\int_V \mathbf{J}_\mathrm{H}^*(\mathbf{r}) \cdot \mathbf{E}^{\mathrm{inc}}(\mathbf{r})\,dv\right\} \\
    &= \frac{1}{2}\mathrm{Re}\{\bar{I}^{\dagger} \bar{V}^{\mathrm{inc}}\}.
\end{aligned}
\label{eq:simulated_pext}
\end{equation}
For both excitations, $P^\mathrm{ext}$ is normalized by its maximum value and denoted by $\tilde{P}^\mathrm{ext}$.

For plane wave excitation, the incident electric field $\mathbf{E}^{\mathrm{inc}}(\mathbf{r})$ is
\begin{equation}
    \mathbf{E}^{\mathrm{inc}}(\mathbf{r}) = \hat{\mathbf{p}} E_0 e^{-\mathrm{j}k_0 \hat{\mathbf{k}}\cdot \mathbf{r}}
\end{equation}
where $\hat{\mathbf{p}}$ is the polarization unit vector, $\hat{\mathbf{k}}$ is the propagation direction, and $E_0=1.0\,\mathrm{V/m}$ is the amplitude of the incident electric field. 

For the excitation by an infinitesimal dipole, the incident electric field $\mathbf{E}^{\mathrm{inc}}(\mathbf{r})$ is given by~\cite{balanis2016}
\begin{equation}
    \mathbf{E}^{\mathrm{inc}}(\mathbf{r})=-\mathrm{j} \omega \mathbf{A}(\mathbf{r})+\frac{1}{\mathrm{j}\omega \mu_0 \varepsilon_0} \nabla[\nabla \cdot \mathbf{A}(\mathbf{r})].
\end{equation}
Here, the vector potential $\mathbf{A}(\mathbf{r})$ is 
\begin{equation}
    \mathbf{A}(\mathbf{r})=\hat{\mathbf{a}} \frac{\mu_0 I_0l}{4 \pi |\mathbf{r}-\mathbf{r}_{\mathrm{d}}|} e^{-\mathrm{j} k_0 |\mathbf{r}-\mathbf{r}_{\mathrm{d}}|}
\end{equation}
where $\hat{\mathbf{a}}$ is the direction of the dipole, $I_0l=1\,\mathrm{A\cdot m}$ is its current moment, and $\mathbf{r}_{\mathrm{d}}$ is its location.

\subsection{Nanosphere}
\label{sec:ex_sphere}
In this example, a metallic nanosphere of radius $1\,\mathrm{nm}$ centered at the origin is considered. Two models are compared: the hydrodynamic (nonlocal) model with the material parameters $\omega_\mathrm{p} = 1.25\times 10^{16}\,\mathrm{rad/s}$, $\gamma=1.36 \times 10^{14}\,\mathrm{rad/s}$, and $v_\mathrm{F} = 1.39 \times 10^6\,\mathrm{m/s}$, and the Drude (local) model with the same values for $\omega_\mathrm{p}$ and $\gamma$ but with $\beta=0$. Simulations are carried out over the frequency range $0.5\,\omega_\mathrm{p}\leq \omega \leq 1.15\,\omega_\mathrm{p}$. A tetrahedral mesh with $N=16\,826$ unknowns is used in the frequency range  $0.5\,\omega_\mathrm{p}\leq \omega \leq 1.07\,\omega_\mathrm{p}$, while a finer mesh with $N=105\,233$ unknowns is used at higher frequencies. For the plane wave excitation, $\hat{\mathbf{p}}=\hat{\mathbf{x}}$ and $\hat{\mathbf{k}}=\hat{\mathbf{z}}$, and for the dipole excitation, $\hat{\mathbf{a}}=\hat{\mathbf{z}}$ and $\mathbf{r}_{\mathrm{d}}=(5,0,0)\,\mathrm{nm}$. 

\begin{figure}[t!]
\centering       
\subfigure[]{\includegraphics[width=0.7\columnwidth]{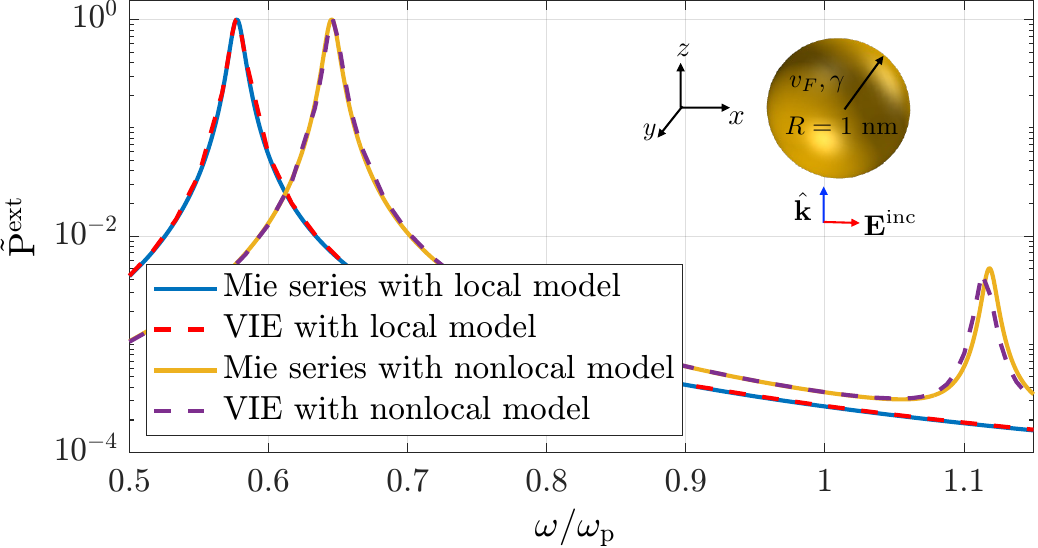}}\\
\subfigure[]{\includegraphics[width=0.7\columnwidth]{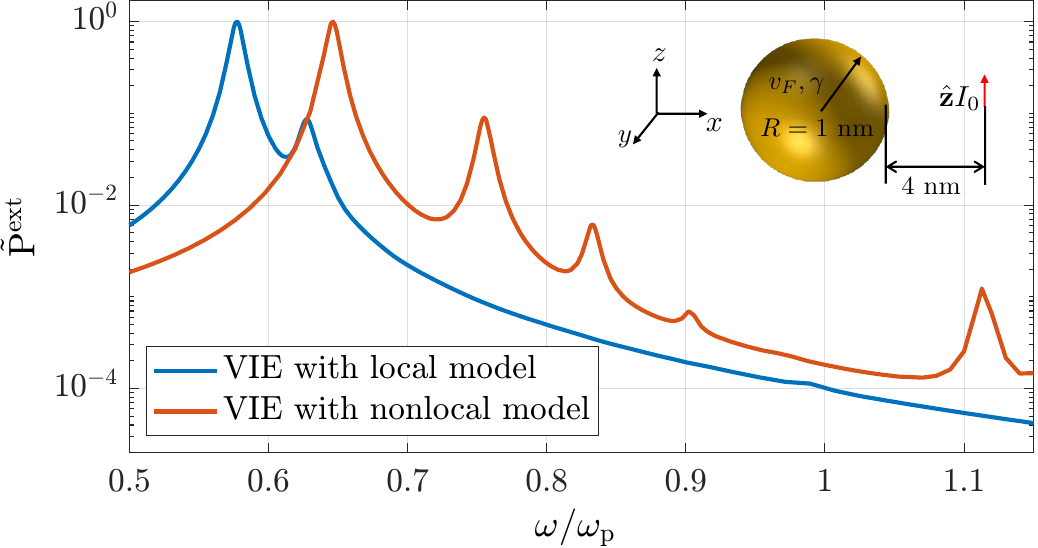}}
\caption{Normalized extinction spectra of the metallic nanosphere under (a) plane wave and (b) dipole excitations.}
\label{sphere_ext}
\end{figure}

Figs.~\ref{sphere_ext}(a) and~\ref{sphere_ext}(b) show $\tilde{P}^{\mathrm{ext}}$ computed using~\eqref{eq:simulated_pext} after $\bar{I}$ is obtained by solving the discretized HDVIE~\eqref{eq:matrix_system} under plane wave and dipole excitations, respectively.

As shown in Fig.~\ref{sphere_ext}(a), $\tilde{P}^{\mathrm{ext}}$ computed with nonlocal and local models under plane wave excitation agrees well with the corresponding Mie series solutions. A small discrepancy appears at higher frequencies in the nonlocal case, which can be reduced by further refining the mesh. Moreover, for the transverse resonances, i.e., those below $\omega_\mathrm{p}$, both excitations exhibit the expected blueshift when nonlocality is included~\cite{doolos2023,raza2011unusual,kupresak2020appropriate}. In addition, the nonlocal $\tilde{P}^{\mathrm{ext}}$ exhibits an extra resonance above $\omega_\mathrm{p}$, at $\omega = 1.13\,\omega_\mathrm{p}$, which is consistent with the longitudinal resonances supported by the hydrodynamic model~\cite{raza2011unusual}. These results highlight the importance of the hydrodynamic model for accurately characterizing metallic nanostructures, particularly near the plasma frequency. Furthermore, in the nonlocal case, $\tilde{P}^{\mathrm{ext}}$ exhibits two resonances at $\omega = 0.65\,\omega_{\mathrm{p}}$ and $\omega = 1.13\,\omega_\mathrm{p}$ under plane wave excitation, whereas five resonances are observed under dipole excitation in the same frequency range. The two resonances excited by the plane wave coincide with two of those excited by the dipole. 

\begin{figure}[t!]
\centering
\noindent
  \includegraphics[width=0.7\columnwidth]{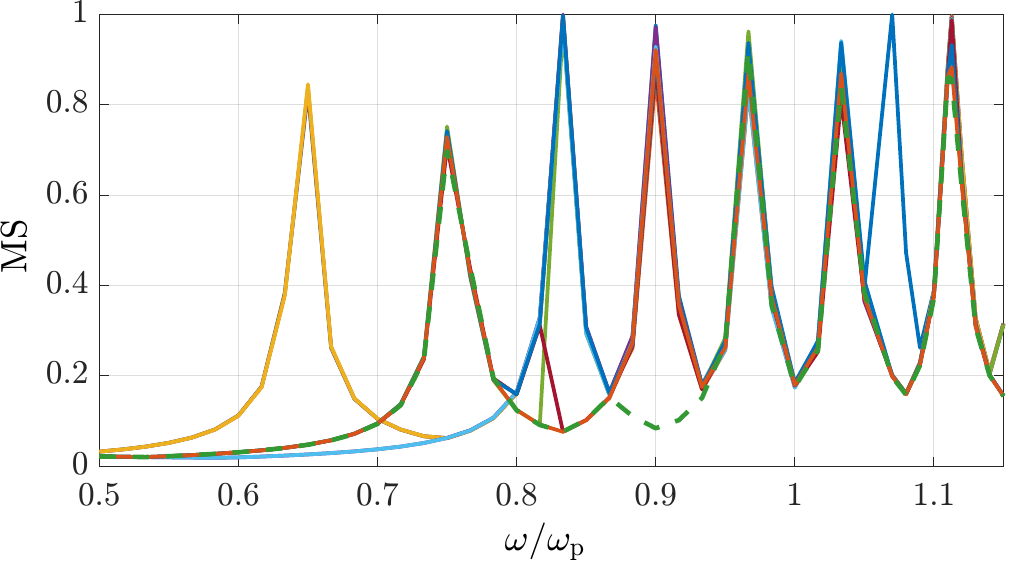}
  \caption{MS curves for the metallic nanosphere.}\label{sphere_ms}
\end{figure}

To identify the intrinsic resonances supported by the structure, the CMA is carried out by solving the GEE in~\eqref{eq:gee2} at $40$ frequency points. The resulting MS curves for the $10$ modes with the largest MS values are shown in Fig.~\ref{sphere_ms}.

\begin{figure}[t]
    \centering
        \subfigure[]{\includegraphics[width=0.7\columnwidth]{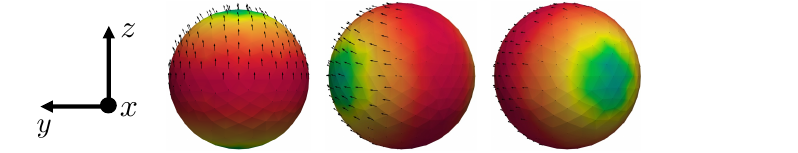}}\\
        \subfigure[]{\includegraphics[width=0.7\columnwidth]{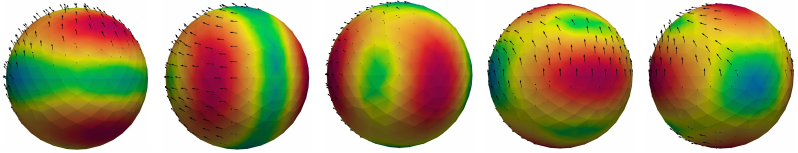}}
    \caption{Visualization of the dominant characteristic mode currents at (a) $\omega = 0.65\,\omega_{\mathrm{p}}$ and (b) $\omega = 0.75\,\omega_{\mathrm{p}}$, showing the current magnitude and direction.}
    \label{fig:vis_sphere}
\end{figure}

Fig.~\ref{sphere_ms} shows that the CMA identifies additional resonances that do not appear as peaks in the extinction spectra, indicating that these modes are not efficiently excited by either the plane wave or dipole source. At the same time, all observed extinction peaks coincide closely with resonances identified by the CMA. While the first resonance ($\omega = 0.65\,\omega_{\mathrm{p}}$) is efficiently excited under plane wave illumination, the second one ($\omega = 0.75\,\omega_{\mathrm{p}}$) is strongly suppressed, as seen in Fig.~\ref{sphere_ext}(a). The dominant characteristic mode currents at these two resonances are shown in Fig.~\ref{fig:vis_sphere}. 

\begin{figure}[t]
    \centering
        \subfigure[]{\includegraphics[width=0.7\columnwidth]{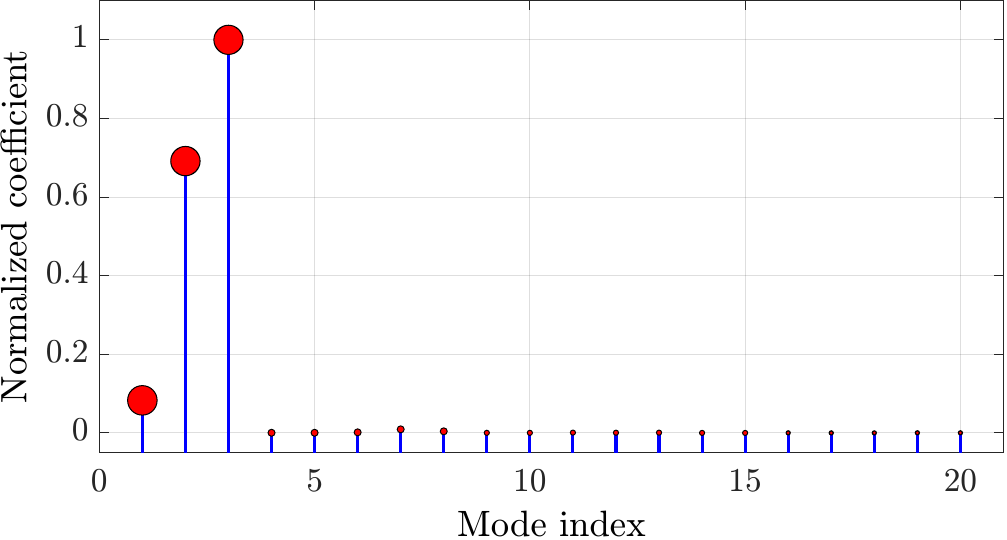}}\\
        \subfigure[]{\includegraphics[width=0.75\columnwidth]{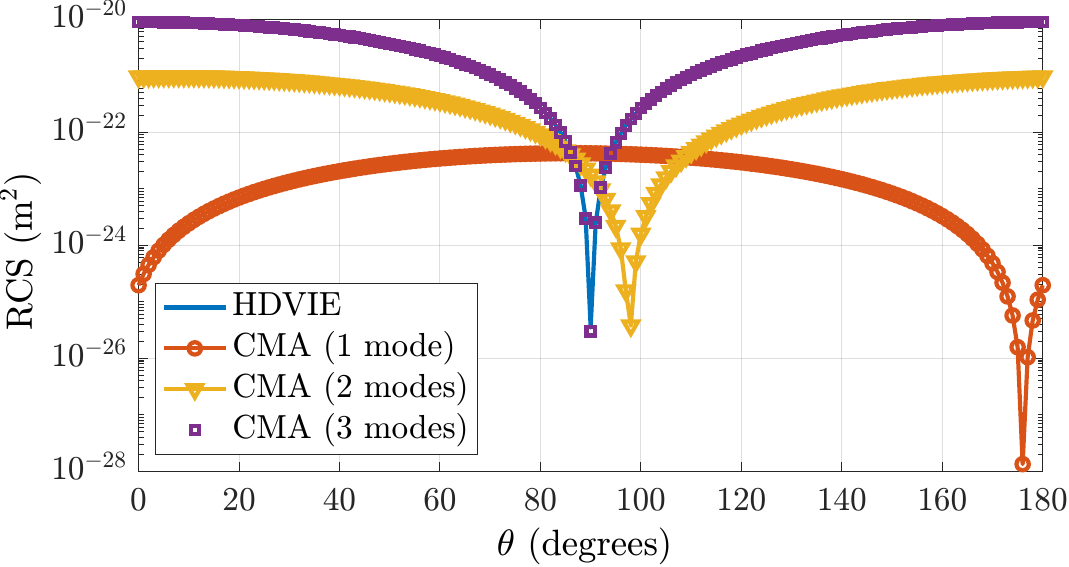}}
    \caption{(a) Normalized modal excitation coefficients, where the marker size represents the MS value. (b) Reconstruction of the RCS from the characteristic modes at $\omega=0.65\,\omega_\mathrm{p}$.}
    \label{fig:rcs_recon_rf1}
\end{figure}

Radar cross section (RCS) is then used to clarify why some resonances identified by the MS curves do not appear as peaks in the extinction spectra. Although these modes are resonant, their contribution to the observed response depends on how strongly they couple to the excitation source. The RCS is computed as
\begin{equation}
    \mathrm{RCS}(\theta, \phi) = \lim_{r\to\infty}  4 \pi r^2 \frac{|\mathbf{E}^{\mathrm{sca}}(\theta, \phi)|^2}{E_0^2}.
\end{equation}
Fig.~\ref{fig:rcs_recon_rf1}(a) shows the normalized modal excitation coefficients $V_k=\bar{I}_k^{\top} \bar{V}^{\mathrm{inc}}$ for the first $20$ modes at the first resonance frequency, $\omega=0.65\,\omega_\mathrm{p}$, where the marker size represents the MS value $\sigma_k$.  

The modes are indexed in descending order of MS. As seen in Fig.~\ref{fig:rcs_recon_rf1}(a), three resonant modes couple strongly to the incident field. Accordingly, the RCS can be reconstructed accurately using the scattered fields associated with only these three modes, as shown in Fig.~\ref{fig:rcs_recon_rf1}(b). The figure shows that the reconstructed RCS matches well with the RCS computed using the solution of the discretized HDVIE~\eqref{eq:matrix_system}.

\begin{figure}[t]
    \centering
        \subfigure[]{\includegraphics[width=0.7\columnwidth]{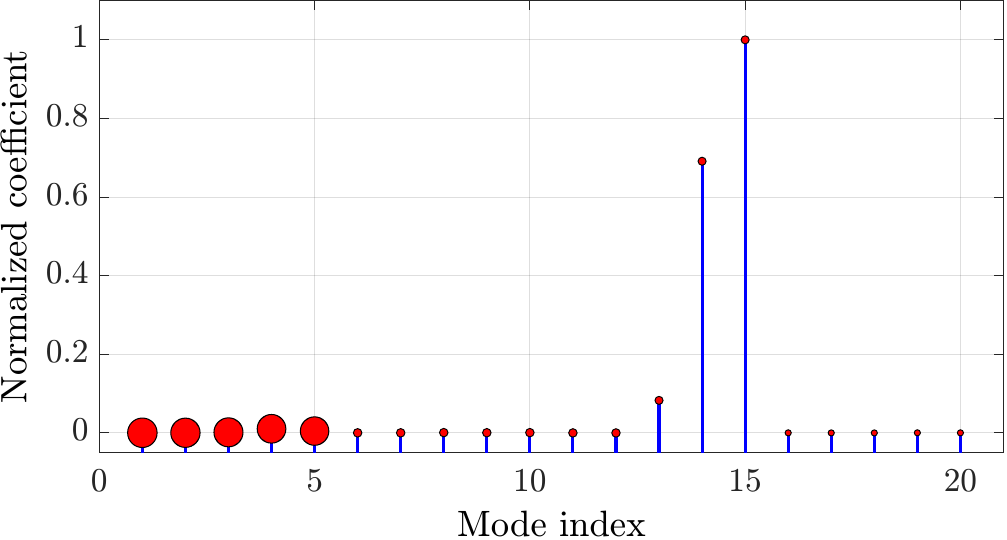}}\\
        \subfigure[]{\includegraphics[width=0.75\columnwidth]{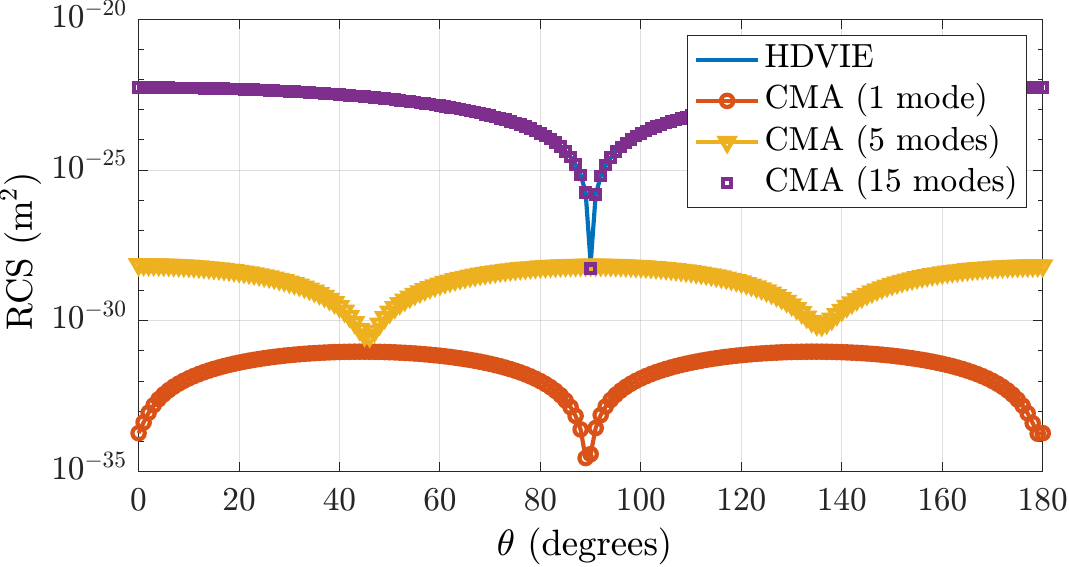}}
    \caption{(a) Normalized modal excitation coefficients, where the marker size represents the MS value. (b) Reconstruction of the RCS from the characteristic modes at $\omega=0.75\,\omega_\mathrm{p}$.} 
    \label{fig:rcs_recon_rf2}
\end{figure}

At the second resonance frequency, $\omega=0.75\,\omega_\mathrm{p}$, Fig.~\ref{fig:rcs_recon_rf2}(a) shows that the dominant modes, associated with high MS values, couple only weakly to the plane wave, which explains the suppression of the corresponding extinction resonance peak.

Instead, stronger coupling is observed for nonresonant modes, specifically modes $13$--$15$. As a result, accurate reconstruction of the RCS at this frequency requires the first $15$ modes, as shown in Fig.~\ref{fig:rcs_recon_rf2}(b).
A comparison of Figs.~\ref{fig:rcs_recon_rf1}(a) and~\ref{fig:rcs_recon_rf2}(a) shows that the three strongly coupled modes have nearly identical excitation coefficients at $\omega=0.65\,\omega_\mathrm{p}$ and $\omega=0.75\,\omega_\mathrm{p}$. Their modal current distributions are also very similar, although not shown here. This indicates that the plane wave couples to essentially the same set of modes at both frequency points. However, these modes are resonant only at $\omega=0.65\,\omega_\mathrm{p}$, whereas they become nonresonant at $\omega=0.75\,\omega_\mathrm{p}$, which explains the absence of a resonance peak. Overall, these results show that the reconstructed RCS converges progressively toward the reference HDVIE result as more modes are included. They also indicate that when resonant modes couple strongly to the incident field, an accurate RCS reconstruction can be obtained using only a small number of dominant modes.
% Discussion of transverse and longitudinal resonances-->

\begin{figure}[t]
    \centering
    \subfigure[]{\includegraphics[width=0.5\columnwidth]{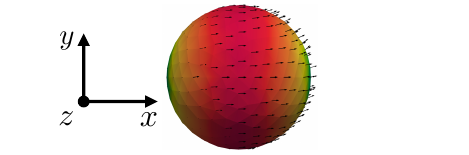}}
    \subfigure[]{\includegraphics[width=0.5\columnwidth]{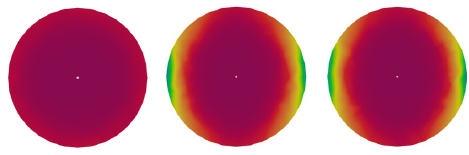}}
    \caption{Visualization of the hydrodynamic current reconstructed from the characteristic modes at $\omega=0.65\,\omega_\mathrm{p}$: (a) current magnitude and direction and (b) current magnitude on the $yz$-, $xz$-, and $xy$-planes.}
    \label{fig:vis_sphere_T}
\end{figure}

\begin{figure}[t]
    \centering
    \subfigure[]{\includegraphics[width=0.5\columnwidth]{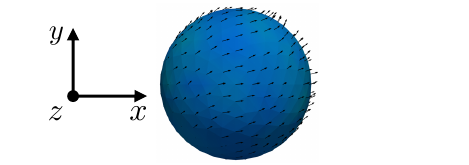}}
    \subfigure[]{\includegraphics[width=0.5\columnwidth]{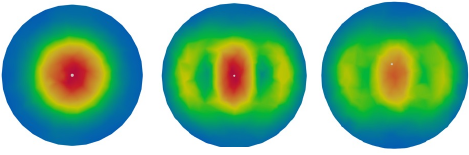}}
    \caption{Visualization of the hydrodynamic current reconstructed from the characteristic modes at $\omega=1.13\,\omega_\mathrm{p}$: (a) current magnitude and direction and (b) current magnitude on the $yz$-, $xz$-, and $xy$-planes.}
    \label{fig:vis_sphere_L}
\end{figure}

Figs.~\ref{fig:vis_sphere_T} and~\ref{fig:vis_sphere_L} show the hydrodynamic currents reconstructed from the computed characteristic modes at the transverse resonance ($\omega=0.65\,\omega_\mathrm{p}$) and the longitudinal resonance ($\omega=1.13\,\omega_\mathrm{p}$), respectively. The number of characteristic modes required for accurate current reconstruction increases significantly with frequency. At the transverse resonance, only three modes are sufficient [see Fig.~\ref{fig:rcs_recon_rf1}(a)], indicating that the response is dominated by a few strongly resonant modes. In contrast, a much larger number of modes is required to obtain a reasonable reconstruction of the longitudinal current. In this example, $212$ modes are needed, indicating richer modal content and a non-negligible contribution from nonresonant modes. The resulting transverse and longitudinal current distributions are also in qualitative agreement with those reported for a nanowire in~\cite{yan2013green}.

\subsection{Nanorod}
In the second example, a metallic nanorod of length $2.5\,\mathrm{nm}$ and radius $1\,\mathrm{nm}$ aligned along the $z$-axis and centered at the origin is considered. The hydrodynamic model parameters are the same as those used in Section~\ref{sec:ex_sphere}. Simulations are carried out over the frequency range $0.5\,\omega_\mathrm{p}\leq \omega \leq 0.9\,\omega_\mathrm{p}$. A tetrahedral mesh with $N=23\,346$ unknowns is used. For the plane wave excitation, $\hat{\mathbf{p}}=\hat{\mathbf{x}}$ and $\hat{\mathbf{k}}=\hat{\mathbf{z}}$ and for the dipole excitation $\hat{\mathbf{a}}=\hat{\mathbf{z}}$ and $\mathbf{r}_{\mathrm{d}}=(5,0,0)\,\mathrm{nm}$. 

\begin{figure}[t]
    \centering
        \subfigure[]{\includegraphics[width=0.7\columnwidth]{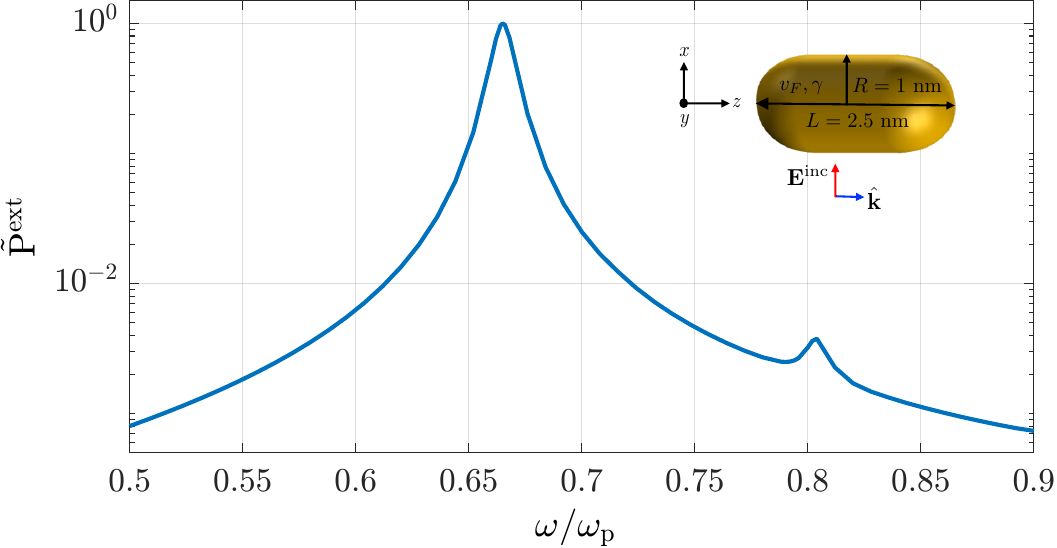}}
        \subfigure[]{\includegraphics[width=0.7\columnwidth]{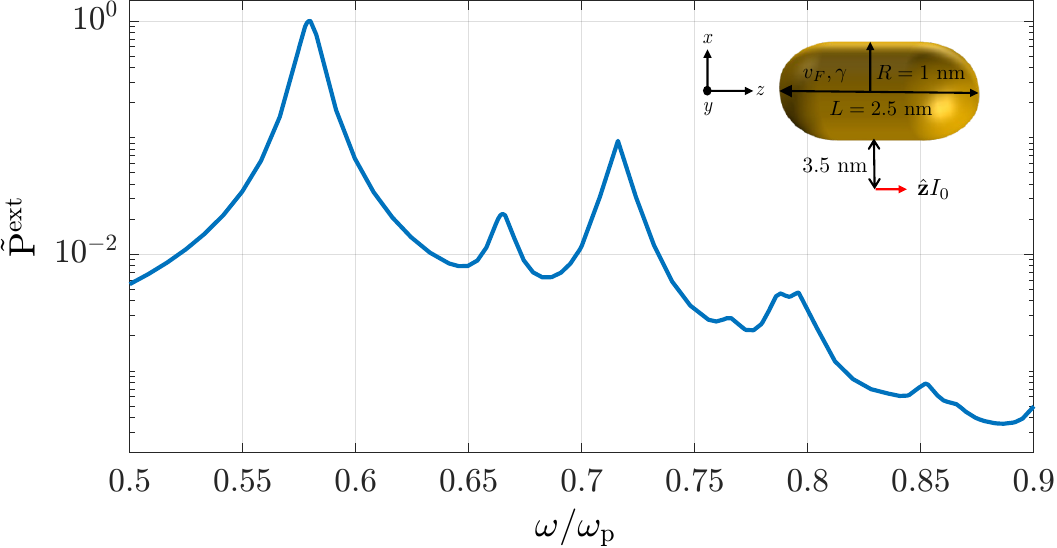}}
    \caption{Normalized extinction spectra of the metallic nanorod under (a) plane wave and (b) dipole excitations.}
    \label{nanorod_ext}
\end{figure}

Figs.~\ref{nanorod_ext}(a) and~\ref{nanorod_ext}(b) show $\tilde{P}^{\mathrm{ext}}$ computed using~\eqref{eq:simulated_pext} after $\bar{I}$ is obtained by solving the discretized HDVIE~\eqref{eq:matrix_system} under plane wave and dipole excitations, respectively. 

The dipole couples to more resonances than the plane wave, although both share a common resonance at $\omega=0.66\,\omega_\mathrm{p}$. 

\begin{figure}[t]
\centering
  \includegraphics[width=0.7\columnwidth]{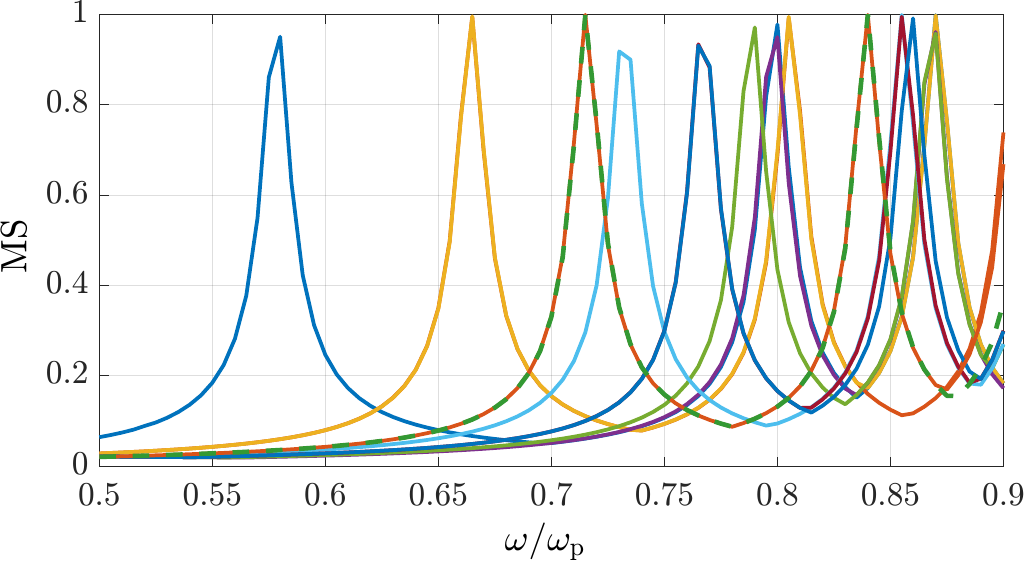}
  \caption{MS curves for the metallic nanorod.}\label{nanorod_ms}
\end{figure}

\begin{figure}[t]
    \centering
    \subfigure[]{\includegraphics[width=0.5\columnwidth]{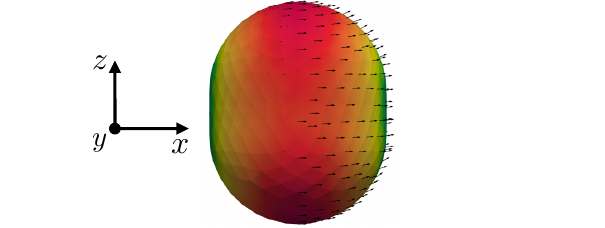}}
    \subfigure[]{\includegraphics[width=0.5\columnwidth]{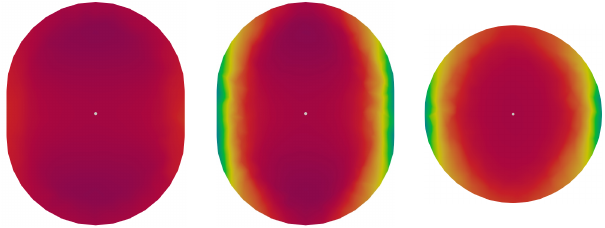}}
    \caption{Visualization of the hydrodynamic current reconstructed from the characteristic modes at $\omega=0.66\,\omega_\mathrm{p}$: (a) current magnitude and direction and (b) current magnitude on the $yz$-, $xz$-, and $xy$-planes.}
    \label{fig:vis_nanorod}
\end{figure}

The CMA is carried out by solving the GEE in~\eqref{eq:gee2} at $81$ frequency points. The resulting MS curves for the $10$ modes with the largest MS values are shown in Fig.~\ref{nanorod_ms}. As in the previous example, the CMA identifies more resonances than the sources excite. Whether a given resonance appears in the extinction spectra depends on how strongly the incident field couples to the corresponding modes. The hydrodynamic current reconstructed from the computed characteristic modes at $\omega=0.66\,\omega_\mathrm{p}$, as shown in Fig.~\ref{fig:vis_nanorod}, exhibits a dipole-like distribution, confirming that this resonance is transverse.

\subsection{Nanodimer}
In the last example, a metallic nanodimer consisting of two spheres of radius $1\,\mathrm{nm}$, with a minimum separation of $0.2\,\mathrm{nm}$ is considered. The dimer is aligned along the $x$-axis and centered at the origin. The hydrodynamic model parameters are the same as those used in Section~\ref{sec:ex_sphere}. Simulations are carried out over the frequency range $0.5\,\omega_\mathrm{p}\leq \omega \leq 0.9\,\omega_\mathrm{p}$. A tetrahedral mesh with $N=33\,388$ unknowns is used. For the plane wave excitation, $\hat{\mathbf{p}}=\hat{\mathbf{x}}$ and $\hat{\mathbf{k}}=\hat{\mathbf{z}}$ and for the dipole excitation $\hat{\mathbf{a}}=\hat{\mathbf{z}}$ and $\mathbf{r}_{\mathrm{d}}=(4.5,0,0)\,\mathrm{nm}$.

\begin{figure}[t]
    \centering
    \subfigure[]{\includegraphics[width=0.7\columnwidth]{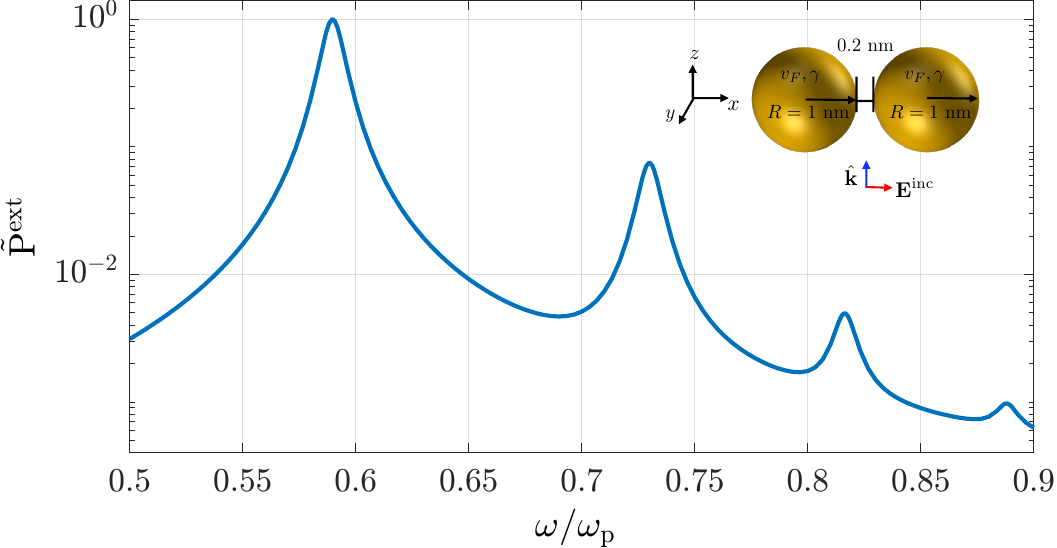}}
    \subfigure[]{\includegraphics[width=0.7\columnwidth]{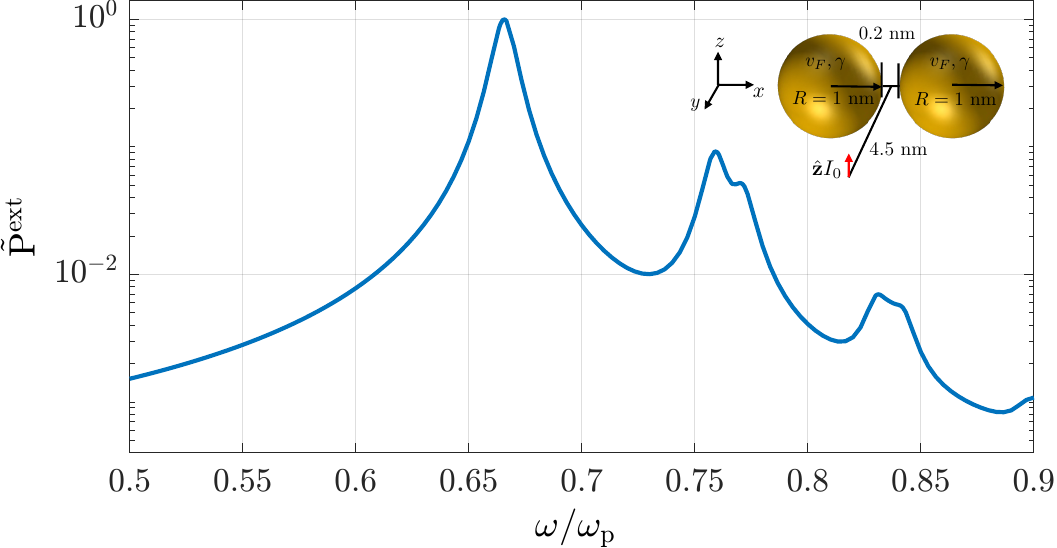}}
    \caption{Normalized extinction spectra of the metallic nanodimer under (a) plane wave and (b) dipole excitations.}
    \label{dimer_ext}
\end{figure}

Figs.~\ref{dimer_ext}(a) and (b) show $\tilde{P}^{\mathrm{ext}}$ computed using~\eqref{eq:simulated_pext} after $\bar{I}$ is obtained by solving the discretized HDVIE~\eqref{eq:matrix_system} under plane wave and dipole excitations, respectively. Unlike the previous examples, the resonances appearing in the extinction spectra under these two excitations do not coincide, since they couple to different subsets of the characteristic modes of the nanodimer. 

\begin{figure}[t]
\centering
  \includegraphics[width=0.7\columnwidth]{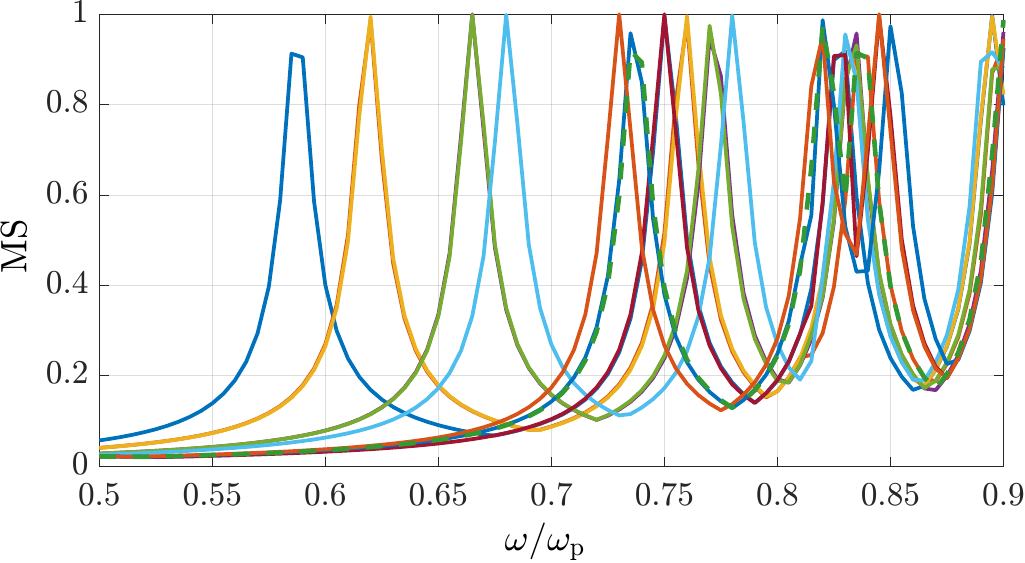}
  \caption{MS curves for the metallic nanodimer.}\label{dimer_ms}
\end{figure}

\begin{figure}[t]
    \centering
    \subfigure[]{\includegraphics[width=0.9\columnwidth]{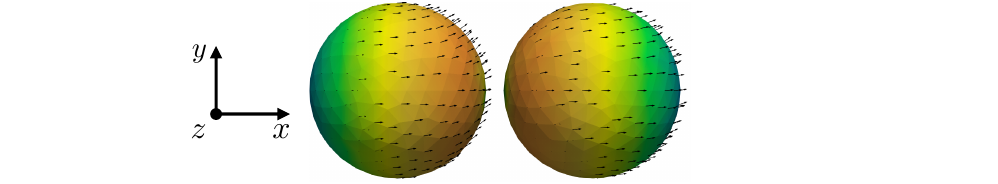}}
    \subfigure[]{\includegraphics[width=0.9\columnwidth]{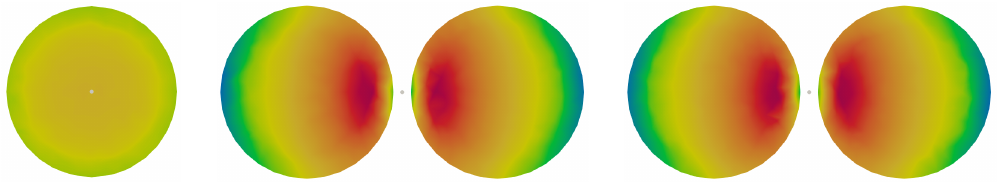}}
    \caption{Visualization of the hydrodynamic current reconstructed from the characteristic modes at $\omega=0.59\,\omega_\mathrm{p}$: (a) current magnitude and direction and (b) current magnitude on the $yz$-, $xz$-, and $xy$-planes.}
    \label{fig:vis_dimer}
\end{figure}

The CMA is carried out by solving the GEE in~\eqref{eq:gee2} at $81$ frequency points. The resulting MS curves for the $10$ modes with the largest MS values are shown in Fig.~\ref{dimer_ms}. Comparison of the extinction spectra with the MS curves shows that the observed extinction peaks coincide closely with the resonances identified by the CMA. Several high-MS modes do not appear in the extinction spectra, which indicates weak coupling to the chosen sources. These results confirm that modal excitation depends not only on the resonant behavior of a mode, but also on its coupling strength to the incident field.
The reconstructed hydrodynamic current at $\omega=0.59\,\omega_\mathrm{p}$ from the computed characteristic modes is shown in Fig.~\ref{fig:vis_dimer}. A strong concentration of current is observed in the gap region, which indicates pronounced near-field coupling between the two spheres.

It should be noted that the minimum gap in this example is $0.2\,\mathrm{nm}$, which lies in the subnanometer regime. At such small separations, additional quantum effects, such as electron spill-out and tunneling, may become important, and the hard-wall boundary condition adopted in the present hydrodynamic model may no longer provide a fully accurate description~\cite{schmitt2016dgtd}. Nevertheless, this example is included to demonstrate the capability of the proposed HDVIE-based CMA framework in analyzing strongly coupled nanostructures.

\section{Conclusion} \label{sec:conclusion}

In this work, the CMA is extended to plasmonic nanostructures that are modeled using the VIE incorporating the hydrodynamic model. Under the assumption of simple (alkali) metals, the coupled system of the HDE and the VIE is reduced to a single HDVIE in terms of the hydrodynamic current density. The HDVIE is discretized using full SWG basis functions. Half SWG functions are excluded to enforce the additional boundary condition required by the HDE at the interface between the metal and the background medium. Galerkin testing then yields the discretized form of the HDVIE as a matrix system, from which the GEE is constructed within the CMA framework. Solving this eigenvalue problem provides excitation-independent access to the intrinsic resonances of the structure through MS curves and the associated characteristic mode currents. Because these modes are intrinsic to the geometry and material of the structure, the induced current under an arbitrary excitation can be reconstructed as a modal expansion, from which electromagnetic quantities such as the RCS can be evaluated.

The proposed HDVIE--CMA formulation is validated through the analysis of three metallic nanostructures: a nanosphere, a nanorod, and a nanodimer. In these examples, the characteristic modes are compared with the extinction spectra under plane wave and dipole excitations and, where applicable, with analytical Mie-series solutions. The CMA identifies the intrinsic resonances of the structure, including those not excited by a given source, while the appearance of a resonance in the extinction spectrum depends on the coupling strength between the source and the corresponding mode. Consequently, selecting an appropriate excitation scheme is essential when a specific resonant mode is targeted in nanoantenna design. These examples also reveal additional resonances arising from the nonlocal response, such as the longitudinal resonance above the plasma frequency in the nanosphere, which are absent in the local model. Beyond identifying resonances, the framework provides physical insight into the underlying modal mechanisms, making it a practical tool for the analysis and design of plasmonic nanoantennas. 

The present formulation is based on the hard-wall boundary condition and therefore does not account for electron spill-out or tunneling. Incorporating such effects into the proposed framework will be considered in future work, particularly for structures with subnanometer features, where additional quantum effects may become significant. 

\bibliographystyle{unsrt}
\bibliography{references_submission_final}

\newpage\clearpage

\section*{Figures}

\end{document}